\documentclass{optica-article}

\journal{opticajournal} 

\articletype{Research Article}
\usepackage{tablefootnote}
\usepackage{lineno}
\usepackage{gensymb}
\usepackage{physics}
\usepackage{tabularx}

\begin{document}

\title{Continuous Entanglement Distribution from an AlGaAs-on-Insulator Microcomb for Quantum Communications}

\author{Trevor J. Steiner,\authormark{1,*} Maximilian Shen,\authormark{2}, Joshua E. Castro,\authormark{2} John E. Bowers\authormark{1,2}, and Galan Moody\authormark{2}}

\address{\authormark{1}Materials Department, University of California, Santa Barbara, CA 93106\\
\authormark{2}Electrical and Computer Engineering Department, University of California, Santa Barbara, CA 93106\\}

\email{\authormark{*}trevorsteiner@ucsb.edu} 

\begin{abstract*}
Using an aluminum gallium arsenide microring resonator, we demonstrate a bright quantum optical microcomb with $>300$ nm bandwidth and more than 20 sets of time-energy entangled modes, enabling spectral demultiplexing with simple, off-the-shelf commercial telecom components. We report high-rate continuous entanglement distribution for two sets of entangled-photon pair frequency modes exhibiting up to $20$ GHz/mW$^2$ pair generation rate. As an illustrative example of entanglement distribution, we perform a continuous-wave time-bin quantum key distribution protocol with 8 kbps raw key rates while maintaining less than 10$\%$ error rate and sufficient two-photon visibility to ensure security of the channel. When the $>$20 frequency modes are multiplexed, we estimate $>$100 kbps entanglement-based key rates or the creation of a multi-user quantum communications network. The entire system requires less than 110 $\mu$W of on-chip optical power, demonstrating an efficient source of entangled frequency modes for quantum communications. As a proof of principle, a quantum key is distributed across 12 km of deployed fiber on the UCSB campus and used to transmit a 21 kB image with $<9\%$ error.
\end{abstract*}

\date{\today}

\section{Introduction}

The distribution and joint measurement of time-energy entangled-photon pairs is important for applications of quantum technologies such as distributed computing, repeater-based quantum networks, and quantum communication \cite{Mosley2008,Knill2001,Yuan2010,moody20222022}. The generation of entangled photon pairs near the 1.5 $\mu$m telecommunications band is especially desirable for low-loss transmission, allowing for quantum communication networks using established optical fiber systems \cite{Gisin2007}. Photonic integrated circuits (PICs) provide additional benefits for low size, weight, and power (SWaP) quantum light transmitters and receivers \cite{Liu2022}. Numerous chip-scale platforms have been demonstrated for entanglement generation and distribution, many of which operate near the 1.5 $\mu$m telecommunications band \cite{Moody2020}. Of these platforms, one of the most promising is aluminum gallium arsenide-on-insulator (AlGaAsOI), which has recently demonstrated ultra-bright, high-quality entangled-photon pairs from a spontaneous four wave mixing process in a microring resonator \cite{Steiner2020}. Improvements to the fabrication process have enabled low propagation loss $<0.2$ dB/cm and high quality factor microcomb resonators with $Q>3.5$ million \cite{Xie2020}. Other standard photonic components including beamsplitters, filters, interferometers, and detectors have also been demonstrated with comparable performance to their silicon counterparts\cite{Castro2022,McDonald2019}. Finally, the AlGaAsOI platform is naturally compatible with InGaAs quantum dot laser sources and boasts large second and third order nonlinearities, making it an excellent platform for quantum photonic integrated circuits (QPICs) with advanced functionality for the generation, manipulation, and detection of quantum photonic signals. 

\begin{figure}[!tb]
\centering
\includegraphics[width=\columnwidth]{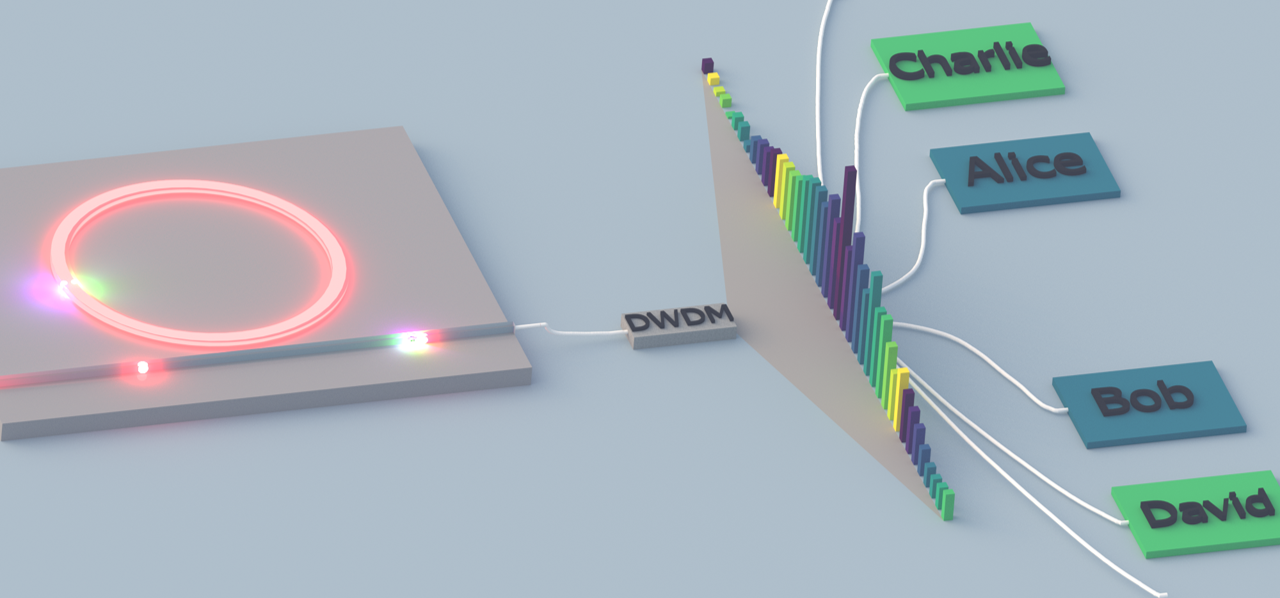}
\caption{AlGaAsOI quantum optical microcomb enabling multi-user entanglement distribution. A microring resonator generates pairs of entangled photons across many spectral modes that can be distributed to various users over long distances using low-loss telecommunication optical fibers. The frequency comb spectrum shown is from the device characterized in this work and shows the ability to create a user network with more than 20 nodes with ultrabright entanglement generation requiring $<100$ $\mu$W on-chip power. In this illustration, Alice and Bob each receive a photon from one set of modes, and Charlie and David each receive a photon from a different set.}
\label{fig:intro}
\end{figure}

One particularly impactful use of entangled-photon pairs generated in the telecom C-band is for quantum key distribution (QKD). QKD is a method to create a secret and random bit string shared between two remote users for secure information distribution\cite{Xu2020}. Since the first proposal of QKD in 1984 by Bennett and Brassard\cite{Bennett2014}, numerous protocols and techniques have been implemented to demonstrate secure key distribution\cite{Sibson2017,Pirandola2020,Kumar2021} including system demonstrations with up to 830 km of deployed optical fiber\cite{Wang2022} and satellite-based QKD exceeding 1,200 km\cite{Liao2017,Bedington2017}. Sources for the secure key are typically entangled photon pairs, single photons, squeezed light, or weak coherent states\cite{Xu2020,Liu2022,Kwek2021}. As source and detector technologies have advanced in recent years, there has been a push for low-SWaP approaches relying on chip-scale implementations that take advantage of existing semiconductor and photonic manufacturing techniques (see reference \cite{Liu2022} for a summary of recent advancements in chip-based QKD technologies). Here we focus on entanglement-based QKD which is inherently secure due to the quantum no-cloning theorem and the collapse of the quantum state upon measurement\cite{Wootters1982,Xu2020}. For entanglement-based protocols, two photons are generated simultaneously through a nonlinear optical process such as spontaneous parametric down conversion (SPDC) or spontaneous four wave mixing (SFWM), and each photon is sent to either Alice or Bob for measurement to establish a key\cite{Diamanti2016}.

\begin{figure}[!tb]
\centering
\includegraphics[width=\columnwidth]{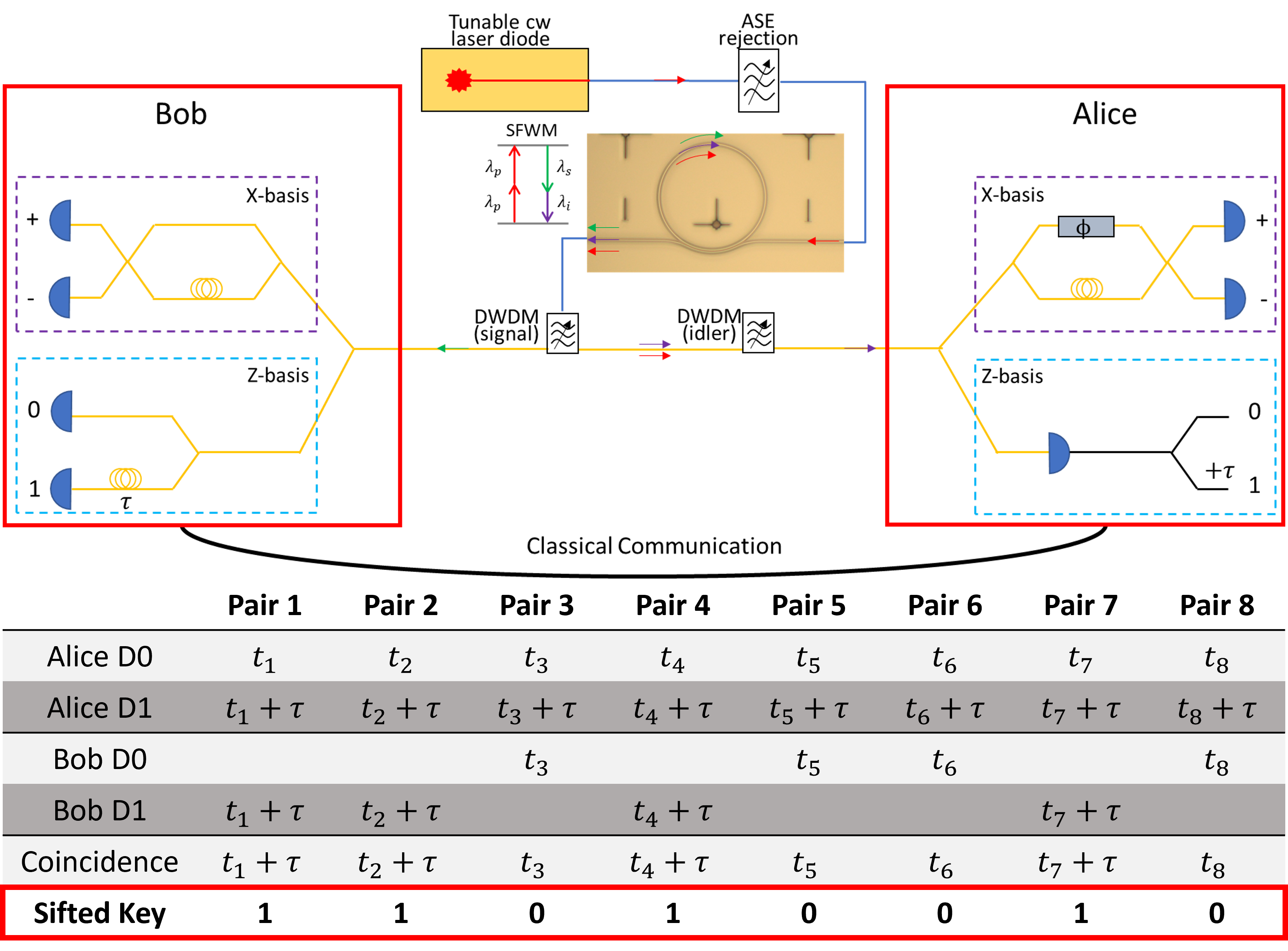}
\caption{Experimental design for a three-detector time-bin QKD protocol. A tunable continuous wave laser source pumps the resonator source after passing through amplified spontaneous emission filters. The pump light, along with the generated entangled photon pairs couple from the chip to lensed optical fibers and enter dense wavelength division demultiplexer filters. The signal photon travels to Bob's measurement setup, and the idler photon travels to Alice. Alice and Bob establish visibility measurements in the X basis via locked Franson interferometry. Alice has a tunable phase shifter and sweeps the relative phase between the short and long path to show two-photon interference of the  $\frac{1}{\sqrt{2}}\left(\ket{SS} + exp(i\phi_{i+s})\ket{LL}\right)$ state. In the Z basis, Alice uses a single superconducting nanowire single photon detector (SNSPD) and collects the single photon arrival times as well as an electronically delayed channel (black lines). Bob splits his signal photons into two paths using a fiber-based 3 dB splitter. Each path has an SNSPD, and Bob records detection events in either the "0" time bin (short path) or "1" time bin (long path). The table shows how a sifted key is established.}
\label{fig:protocol}
\end{figure}

In this work, we generate entangled-photon pairs generated from spontaneous four-wave mixing (SFWM) in an AlGaAsOI microring resonator \cite{Steiner2020} in conjunction with a low-loss photonic testbed consisting of double-pass dense wavelength division multiplexers (DWDMs from Fiberdyne Labs) to demonstrate a three-detector, continuously pumped, time-bin QKD protocol \cite{Pelet2023}. The AlGaAsOI resonator generates a broad spectrum spanning $>300$ nm of entangled-photon pairs with larger free-spectral range for simple DWDM, which can be used to either spectrally multiplex for higher secure key distribution rates or used to send different pairs to numerous users, creating a multi-user connected quantum information network as demonstrated in Fig. \ref{fig:intro}. We demonstrate entanglement-based QKD using two sets of time-bin entangled modes from the comb. We record raw key rates and raw key efficiencies up to 8 kbps and 74 kbps/mW of pump power, respectively, for a single pair. The efficiency, which surpasses the record efficiency from bulk PPLN \cite{Neumann2022_1Gb}, is enabled by the tight optical confinement in the highly nonlinear and low-loss microring resonator, demonstrating the advantage of PICs for quantum communication applications that require low-SWaP.

\subsection*{Protocol}\label{sec:protocol}
We follow a recently developed three-detector asymmetric sifting QKD protocol\cite{Pelet2023} that does not require a pulsed laser for clock synchronization. The protocol also enhances the secure key rate by a factor of 2 by requiring one fewer detector than similar implementations\cite{Neumann2022}, which enabled 7 kbps raw key rates using a bulk SPDC source\cite{Pelet2023}. Here, we expand on this demonstration through the use of a SFWM chip-scale source that reduces the input optical power while maintaining high key rates. 

The protocol is shown in Fig. \ref{fig:protocol}. A tunable continuous wave (cw) laser source is sent through amplified spontaneous emission (ASE) filters to remove any pump power at the adjacent signal and idler frequencies used for the protocol. The pump light is coupled into/out of an AlGaAsOI ring resonator with 1 THz free-spectral range using lensed optical fibers, whereby entangled-photon pairs are generated in spectral modes adjacent to the pump through SFWM. In Fig. \ref{fig:protocol}, the inset shows the energy conservation of the SFWM process, where $\lambda_p$, $\lambda_i$, and $\lambda_s$ correspond to the pump, idler, and signal photons\cite{Helt2010}. Once the photon pair is generated, the signal and idler photon are demultiplexed using double-pass 100 GHz DWDMs to provide sufficient extinction of the pump light and to separate the signal and idler photons into respective fiber channels. The signal photons are sent to Bob while the idler photons are sent to Alice. Both Alice and Bob utilize a beamsplitter to send their stream of photons to either the X or the Z basis; the Z basis is used to generate the key and measures the photon arrival time in one of two time bins separated by a user-defined, fixed delay $\tau$. The X basis monitors the signal-idler two-photon interference visibility to determine the security of the key, where a visibility $>50\%$ surpasses the classical limit and $>70.7\%$ violates the Clauser-Horne visibility \cite{Clauser1974,Rarity1990}. .

A key innovation for cw time-bin QKD introduced by Pelet \textit{et al.} is that Alice and Bob have unique Z basis measurement systems \cite{Pelet2023}. The table in Fig. \ref{fig:protocol} shows an example of 8 sets of photon pairs arriving to Alice and Bob's Z bases. Bob's Z basis measurement consists of a 3-dB beamsplitter to randomly send the signal photons along one of two optical paths corresponding to a 0-bit (short optical path with time $t_n$ in the table in Fig. \ref{fig:protocol}) or a 1-bit (long optical path with time $t_n + \tau$). Bob uses two superconducting nanowire single photon detectors (SNSPDs) to measure the photon's arrival time within 100 ps time-bins and $>85\%$ detector efficiency. Contrary to Bob, Alice utilizes a \textit{single} SNSPD to detect the arrival time of the idler photons. With her arrival times, she creates a second, electronically delayed channel with a delay equal to Bob's long fiber path delay, $\tau$. Alice thus records a detection event at $t_n$ and $t_n$+$\tau$ for each detected idler photon from the source. Using a classical channel, Bob only communicates to Alice the arrival time of photons at his detector     without sharing any path information. By comparing the arrival times and looking for zero-time delay coincidences, Alice is able to determine whether Bob detected the photon in his 0- or 1-bit path. Through this post-selection, Alice and Bob establish their sifted key based on the photon coincidence with Bob's 0-bit or 1-bit detector. The table in Fig. \ref{fig:protocol} illustrates the protocol with eight exemplary measurement outcomes and the corresponding sifted key. Common time-bin protocols utilize a pulsed laser for synchronization and a 3-dB splitter on both Alice and Bob's Z basis measurement, which limits the secure key rate to 25$\%$ of the generated coincidence count rate\cite{Neumann2022}. Here, the raw coincidence count rate usage is improved to 50$\%$ by removing the choice of two paths on Alice's side.

The X basis monitors the entanglement visibility via Franson interferometry \cite{Franson1989}. Each photon enters an unbalanced Mach-Zehnder interferometer separating the photon coincidences into three distinct time bins: an early bin when the idler takes the short path and the signal takes the long path, a late time bin when the idler takes the long path and the signal takes the short path, and a central bin when both photons take the same path (see Fig. S2). This central bin corresponds to the photon state: $\frac{1}{\sqrt{2}}\left(\ket{SS} + exp(i\phi_{i+s})\ket{LL}\right)$ where the phase $\phi_{i+s}$ is set by a tunable phase shifter in the short arm. Alice's and Bob's interferometers are locked to maintain a consistent relative phase during the measurement, and Alice's voltage controlled phase shifter can be swept to change the phase between the two states. Importantly, because the time bins are defined at Alice and Bob, only the interferometers at Alice and Bob need to be phase stable, whereas phase fluctuations in the fibers from the source to Alice and Bob do not impact the phase of the time-bin entangled pairs defined once the photons arrive at the nodes.

\section{Source Characterization}

\begin{figure}[!b]
\centering
\includegraphics[width=\columnwidth]{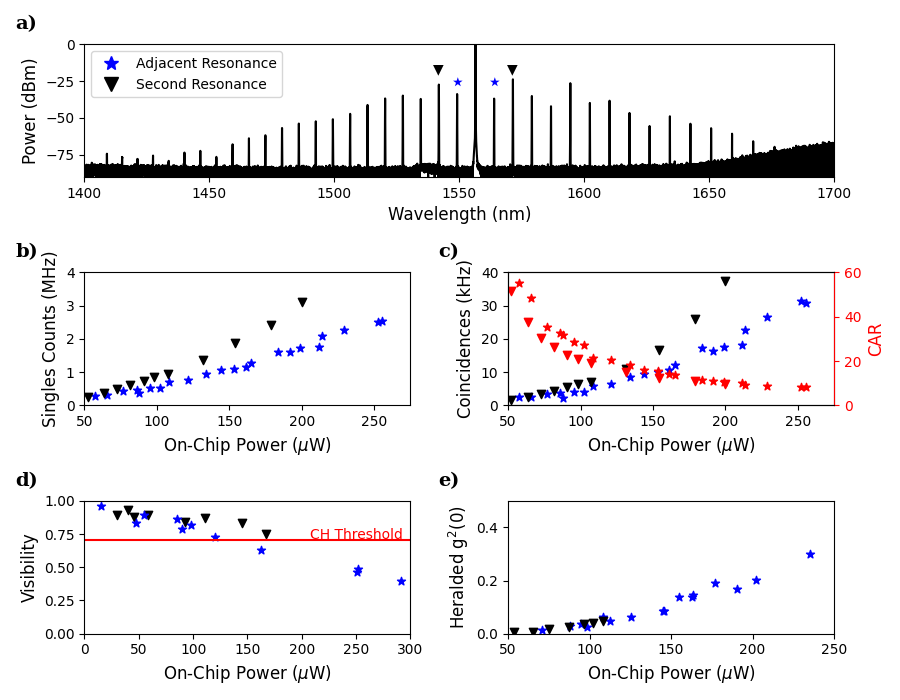}
\caption{Characterization of the entangled photon pair source used for the QKD protocol. a) The frequency comb spectrum from the ring resonator source when pumped with sufficient pump power to generate a frequency comb visible using an optical spectrum analyzer. The blue stars indicate the adjacent resonance and the black triangles indicate the second nearest resonance. b) Collected raw single photon counts as a function of the on-chip pump power. c) Measured coincidence count rate as a function of on-chip power along with the coincidence-to-accidental ratio (CAR). d) Raw (without background subtraction) two-photon interference visibility as a function of the on-chip power. The visibility is largely limited by the pump leakage at higher powers. Fig. S3 in the Supplemental shows the visibility after background subtraction. e) Heralded single photon purity measurement as a function of power. }
\label{fig:source}
\end{figure}

The source used to generate entangled photon pairs is an AlGaAsOI microring resonator (a microscope image of a comparable source is shown in Fig. \ref{fig:protocol}). The microring resonator studied in the following experiment had a radius of 13.91 $\mu$m. The width of the bus waveguide was 0.48 $\mu$m, and the ring waveguide was 0.69 $\mu$m wide. The gap between the waveguide and the ring was 0.48 $\mu$m, and the AlGaAs layer was 0.4 $\mu$m thick. For full device fabrication details as well as additional device characterization, refer to our previous work \cite{Steiner2020}. The resonator is pumped at the resonance near 1557 nm to generate a quantum frequency comb that spans over 300 nm. When pumped with sufficient power (above the optical parametric oscillator threshold), the frequency comb is visible with an optical spectrum analyzer showing the possible generation of >20 pairs of entangled photon pair frequencies, as shown in Fig. \ref{fig:source}a). Above 1650 nm, the optical spectrum analyzer has a limited responsivity and thus exhibits an increasing background that eventually surpasses the generated comb lines. In Fig. \ref{fig:source}a), the adjacent and second nearest resonances are denoted by the blue star and black triangle, respectively. Throughout this work, data collected from both resonances will be presented. Data from the adjacent resonance will be plotted using blue stars, and data from the second nearest resonance will be black triangles. As described in the previous section, the QKD protocol relies on a single pair of photon modes, thus such a large bandwidth can be used for either wavelength multiplexing (enhancing the key rate by a factor of $\sim20$) or for multi-user distribution of quantum information. 

First, we bypass the beamsplitters that distribute photons to the X and Z basis shown in Fig. \ref{fig:protocol} and send the signal and idler photons directly to the SNSPDs to characterize the raw single photon count rate and coincidence count rates. The low-loss DWDM filters ($<$3 dB insertion loss with $>$100 dB extinction) and optimized chip-to-fiber coupling ($\sim$3.5 dB per facet) enable efficient collection of signal and idler photons with high coincidence count rates for low pump powers. The pump resonance (near 1557 nm) has a loaded quality factor $Q = 1.24$ million while the adjacent resonances (blue stars) have quality factors of 0.67 and 0.39 million, and the second nearest resonances have quality factors of 1.05 million and 1.00 million (black triangles). The lower quality factors for the nearest resonance modes reduces the entangled-photon pair generation rate, however, even with the nearest modes, over 2.5 MHz detected single photon counts and 30 kHz coincidence counts are recorded (Fig. \ref{fig:source}b,c)) with less than 300 $\mu$W of on-chip pump power. The coincidence to accidental ratio (CAR) of the source remains above 8 at all powers, indicating a high quality, low loss photonic testbed capable of generating a secure key with low errors and high rates. For the second nearest resonance, we measure up to 40 kHz raw coincidences and a measured singles count rate above 4 MHz, which is limited by latching of the SNSPDs.

The security of the protocol relies on monitoring the entanglement visibility of the source during operation. Fig. \ref{fig:source}d) shows the visibility of the source as a function of on-chip power without background subtraction. Here, the visibility is monitored using a folded Franson interferometer where both the signal and idler photon enter the same interferometer before being demultiplexed from each other. In a field-deployed version of the quantum key distribution system, Alice and Bob would each have their own interferometer with a calibrated relative phase. As an example, the visibility of the second resonance mode using 170 $\mu$W of on-chip power is shown in Fig. S2. At higher powers, the extinction of our filters is not sufficient to block all of the pump light from reaching the detectors, which degrades the two-photon visibility. As a result, above 120 $\mu$W of on-chip power, the visibility is below the Clauser-Horne (CH) threshold\cite{Clauser1974}. We thus keep the on-chip pump power below 140 $\mu$W during the QKD protocol. The second resonance has greater pump extinction since the modes are 2 THz from the pump mode, allowing for a higher pump power to be used while maintaining sufficient visibility; however, to keep the data compatible with multiplexed QKD systems, the pump power is limited to 120 $\mu$W to be well above the CH threshold for both the adjacent and second resonance modes. In the supplemental, the visibility with background subtraction is reported (Fig. S3), indicating that with additional pump filtering, up to 275 $\mu$W of pump power can be used while still remaining above the CH threshold. At large enough powers, the multi-photon pair emission probability begins to contribute considerably, opening up the potential risk of photon-number splitting attacks on the QKD system \cite{Sun2022}.

Another important property of the source is the heralded single-photon purity, which can be used to quantify the probability of generating multiple photon pairs in a given time bin. This characterization is completed using a heralded g$^2\left(0\right)$ measurement where the signal photons are detected directly, and the idler photons are sent to a 3-dB splitter before being detected at one of two detectors. Since the photon pairs should be generated simultaneously, recording the three-fold coincidences between the detectors should show a dip at zero time delay. The value of the heralded g$^{(2)}_H\left(0\right)=\frac{N_{ABC}N_{A}}{N_{AB}N_{AC}}$ where $N_{ABC}$ is the number of threefold coincidences, $N_{A}$ is the heralding count rate, and $N_{AB,AC}$ are the twofold coincidences between the heralding channel and one of the two idler channels. An exemplary scan of the heralded g$^2\left(0\right)$ for the source described here is shown in the supplemental (Fig. S1). Fig. \ref{fig:source}e) shows the heralded g$^2\left(0\right)$ as a function of on-chip power. For both sets of signal and idler modes, the single photon purity remains above 90$\%$ when the on-chip power is below 150 $\mu$W.

\section{Time Bin Quantum Key Distribution}
\begin{figure}[!htb]
\centering
\includegraphics[width=\columnwidth]{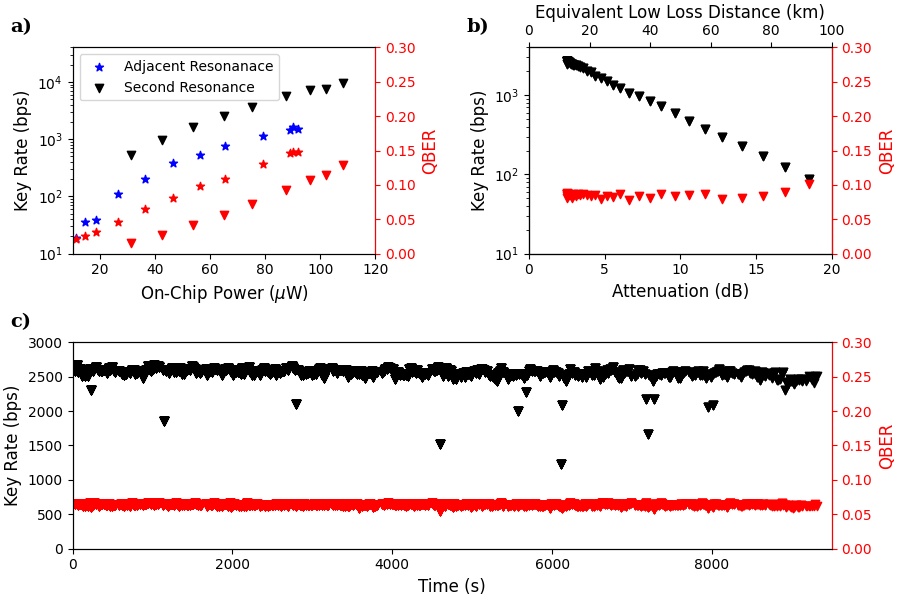}
\caption{a) Raw key rate and bit error rate as a function of on-chip pump power. The pump power is limited to $<$120 $\mu$W to maintain a visibility above 70.7$\%$. The triangles indicate the adjacent resonance characterization, and the stars indicate the second nearest resonance. b) Raw key rate and bit error rate of the second nearest resonance as a function of attenuation on Alice's path. The pump power is maintained at $\sim$90 $\mu$W while sweeping the voltage on a variable optical attenuator. The equivalent optical fiber distance is calculated and indicated on the top axis. c) Key rate stability measurement. Pumping with $\sim$65 $\mu$W of on-chip power, the source consistently outputs 2.5 kbps raw key rate for up to 10,000 seconds (limited by the measurement time of the experiment).}
\label{fig:qkd}
\end{figure}

With a bright source of entangled-photon pairs operating in a regime with high visibility and high single-photon purity, the QKD protocol described in Section \ref{sec:protocol} can be implemented. We monitor the Z basis (shown in Fig. \ref{fig:protocol}) to generate a raw key at various pump powers to assess how the error rates and raw key rates scale with power. Fig. \ref{fig:qkd}a) shows the raw key rates collected for the adjacent and second nearest resonance of the ring resonator source. As described in the source characterization section, the difference in resonance quality factors between these two pairs of modes creates a difference in the generated entangled photon pair flux, and thus the key rate of the second resonance is $\sim$4 times larger than that of the adjacent resonance. Additionally, the pump rejection filters used in this experiment have larger extinction ratios for larger detuning from the pump wavelength, so the error rates of the adjacent resonance are larger than the second nearest resonance due to pump leakage into the detectors. Fig. \ref{fig:source}d) shows the visibility measurement (X basis) for the resonances at different pump powers. Since the Clauser-Horne threshold is surpassed for powers less than 126 $\mu$W, we demonstrate the QKD protocol up to these powers. Without any error correction, raw key rates with less than 10$\%$ error rates of up to 0.6 kpbs and 8 kbps are recorded for the adjacent and second nearest resonance, respectively. A binary secure key can be established with up to $\sim11\%$ error rates at which the Shannon rate $1-2H(x)$ reaches 0 (where $H(x)=-xlog_2(x)-(1-x)log_2(1-x)$ is the binary entropy function)\cite{Neumann2021,Ma2007,Waks2002,Shor2000}. We report the raw key rates up to this bound to ensure that a secure key can be established. With additional pump filtering for the adjacent resonance, it is expected that the rate can approach 2 kbps. 

A comparison of this result with a subset of other entanglement-based QKD demonstrations is included in Table \ref{tab:QKD_Comparison}, and Reference \cite{Xu2020} has an extensive review of recent experimental works. For comparable error rates, our source does not outperform all of the demonstrations in this table; however, the efficiency of this source is improved, requiring only $\sim$100 $\mu$W of optical power to generate a comparable raw key. For example, when normalizing to the pump power, the raw key generation efficiency is $>60$ times higher in this work compared to bulk periodically poled lithium niobate (PPLN) used for the same protocol\cite{Pelet2023}. Through dense spectral multiplexing with $>66$ frequency channels, polarization encoding using bulk PPLN can boost the key rates to $>1$ Gbps\cite{Neumann2022_1Gb} with 400 mW of pump power. Comparing efficiencies for low-SWaP applications such as space-based entanglement distribution, the AlGaAsOI PIC single-pair efficiency of 74 kbps/mW is higher than polarization encoding with bulk PPLN (45 kbps/mW).

\begin{table*}[!ht]
\centering
\resizebox{\columnwidth}{!}{\begin{tabular}{ccccccc}
Platform & Type & Resource & Raw Key Rate (kbps) & Error Rate & Pump Power (mW) & Reference\\
\hline
AlGaAsOI ring & SFWM & Time-bin  & 8& 0.09 &0.108 & This work\\
Bulk PPLN & SPDC & Time-bin  & 7  & 0.047 & $<$6 &\cite{Pelet2023}\\
AlGaAs waveguide & SPDC & Polarization & 0.039 & 0.02 & &\cite{Appas2021}\\
Bulk PPKTP & SPDC & Polarization & 0.109 & 0.064 & 2.4 & \cite{Shi2020}\\
Bulk PPLN & SPDC & Polarization & 0.3 & $<$0.06 & 1-10 & \cite{Joshi2020}\\
Si waveguide & SFWM & Dispersive optics & 0.04-0.06 & 0.07-0.08 & & \cite{Liu2022multi}\\
Bulk PPLN & SPDC & Polarization & $>$100$\times10^3$ & $<$0.07 & 422 & \cite{Neumann2022_248km}\\
Bulk PPLN & SPDC & Polarization & \shortstack{$>$1.8$\times10^4$\\(simulated)} & & 400 & \cite{Neumann2022_1Gb}\\
\end{tabular}}
\caption{\label{tab:QKD_Comparison} Comparison of entanglement-based QKD protocols. The key rates are adjusted to reflect transmission without attenuation and are reported for a single set of photon pairs (i.e without multiplexing).}
\end{table*}




Next, the QKD protocol is extended to artificial distances using a variable optical attenuator placed before Alice's detection setup. Fig. \ref{fig:qkd}b) shows the raw key rate and bit error rate as a function of attenuation. The top axis shows the equivalent distance in kilometers for the attenuation assuming low-loss optical fiber with 0.2 dB/km attenuation. Since the protocol relies simply on the photon arrival times, photon loss will dominate all contributions to the degradation of the raw key rate and increase in error rate. Using the second nearest resonance at an attenuation-free raw key rate of 6 kbps and error rate of approximately 8.5 percent, the attenuator voltage is swept from 2.5 dB of attenuation (the insertion loss of the attenuator) to 18.5 dB of attenuation. The key rate remains above 100 bps for all attenuations, and the error rate remains consistent until approximately 15 dB of attenuation where it gradually increases to $\sim$10 percent at the largest attenuation. With this data, we show an effective key rate of 100 bps at an equivalent distance of 92.5 km with error rates below 10$\%$.

To assess the long-term stability of the source, we show in Fig. \ref{fig:qkd}c) a raw key and error rate measurement for up to 10,000 seconds. The pump laser is held on resonance while monitoring the power through the device. Since the resonance has $\>3$ dB transmission dip, when the device falls off resonance, the monitored power increases by $\>3$ dB, and a rapid wavelength sweep of the pump laser realigns it to the pump resonance. The data is sampled every 20 seconds during the stability measurement, and the data points with key rates lower than the others are caused by this realignment process. Other than this monitoring, the operation requires no additional adjustments to the source. The slight reduction in the key rate over time is caused by fiber drift in the experimental setup, which can be mitigated through fiber packaging. Throughout the 10,000 seconds, the error rate is stable at $\sim$6 percent.

\section{Discussion}

\begin{figure}[!htb]
\centering
\includegraphics[width=\columnwidth]{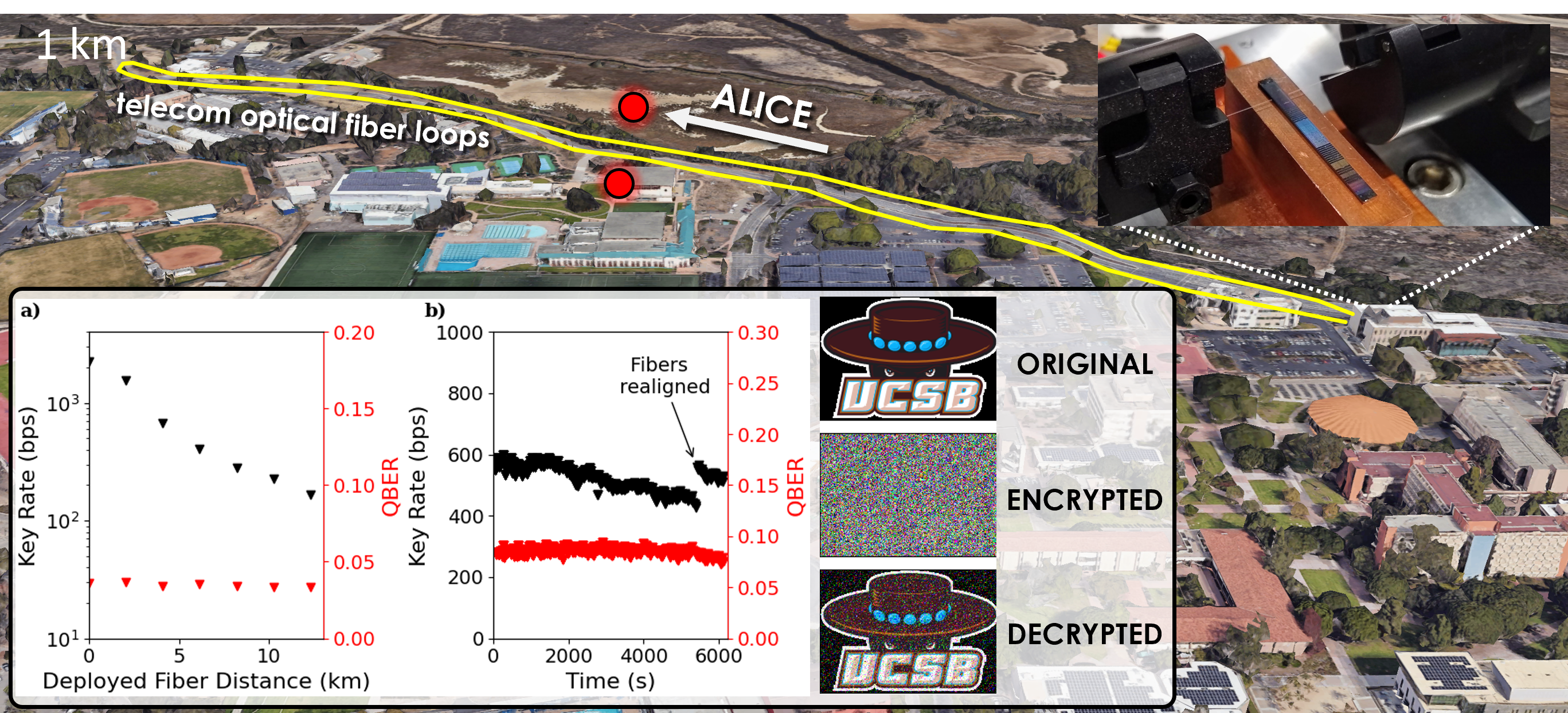}
\caption{Illustration of the deployed fiber experiment conducted on the UCSB campus. Six loops of $\sim$2 km fibers are in place between our research lab in Henley Hall and the campus police station. The yellow lines indicate a rough estimate of the path of the deployed fibers which follow underneath the road connecting these parts of campus. The six independent loops allow for measurements to be completed at different deployed fiber lengths. a) The raw key rate and error rate of the QKD protocol after sending Alice's photons through the deployed fiber path. b) The stability of the key rate and error rate at the longest deployed fiber distance of 12.3 km. The chip-to-fiber coupling drift is significant during this measurement, so a realignment is completed at the 5,300 second mark. Using the 12.3 km deployed fiber, the UCSB logo is transmitted and encrypted using the QKD protocol. The image is transmitted at 600 bps and a 9 percent error rate without any error correction.}
\label{fig:realfiber}
\end{figure}

To demonstrate the practical utility of the source, we send Alice's photon through a deployed, standard SMF-28 telecom fiber on the University of California Santa Barbara campus. The image in the background of Fig. \ref{fig:realfiber} shows an illustration of the approximate deployed fiber path with a $\sim$2 km fiber loop connecting our research lab with another building on campus which is used to demonstrate the robustness of the protocol to external environmental factors. Six fiber loops are made, allowing for $\>12$ km transmission using real fiber lines. Fig. \ref{fig:realfiber}a) shows the key rate and errors for the second nearest resonance mode with an initial rate of 1 kbps and 3 percent error. The deployed fibers exhibit additional loss due to connectors instead of fiber splices being used to configure the number of fiber loops in the channel, which degrades the secure key rate. However, the error rate remains consistent and low for all fiber loop configurations indicating that the influence of environmental factors is negligible. Additionally, the visibility is monitored and is $>73.9\%$ at 12.3 km indicating that the state is not impacted through the long distance distribution.

The stability of the system is also monitored for 6,000 seconds to confirm that external factors like vibrations from vehicles on the road above the fiber or temperature fluctuations do not influence the raw key generation rate stability. As shown in the measurement in Fig. \ref{fig:realfiber}b), fiber drift at the chip-fiber coupling interface is the largest factor in the key rate variation over time. To make sure this variation was from the chip-fiber interface and not a variation in the deployed fiber, a fiber realignment was completed showing that the initial rate can be recaptured and confirming that the deployed fiber does not impact the raw key rate. With this information, we hypothesize that the attenuation-based demonstration accurately models the impact of deployed fiber length on the error rate and key rate. We therefore predict the ability for $>100$ bps key rates with $<10$ percent error at 92.5 km without spectral multiplexing.

Finally, we use the deployed 12.3 km fiber to transmit a key to encrypt and decrypt an RGB image as shown in Fig. \ref{fig:realfiber}. Alice and Bob establish their key at 600 bps with an error rate $<9$ percent (at the same time as the stability measurement shown in Fig. \ref{fig:realfiber}b)) to communicate the UCSB logo. The right side of the inset of Fig. \ref{fig:realfiber} shows the original image, encrypted image and decrypted image using the QKD protocol described in this work. Without error correction, the source shows the capability of transmitting a key across the 12.3 km real deployed fiber to securely encrypt/decrypt a 21 kB image.

\section{Conclusion}
Using a high-quality AlGaAs-on-insulator microring resonator, we generate a quantum optical microcomb spanning $>300$ nm with $>20$ time-energy entangled modes that can be easily demultiplexed and filtered using commercial, off-the-shelf telecom fiber components. We demonstrate the ability to demultiplex photon pairs for scalable entanglement distribution across standard telecom SMF-28 fiber. As an illustrative example, we demonstrate a continuous-wave time-bin quantum key distribution protocol that does not require pulsed optical excitation for clock synchronization. Requiring only $\sim100$ $\mu$W of optical power to generate up to 8 kbps raw key rates with errors $<$10 percent, we demonstrate the low SWaP capability and stability of the source and the protocol using a campus-deployed telecom fiber loop. The source operates with $>$20 frequency mode pairs allowing for key rates exceeding 100 kbps to be achieved through wavelength multiplexing or a connected $>$20-user quantum photonic network to be established.

Looking forward, the low SWaP of the source while maintaining ultra-high entanglement distribution rates opens new opportunities for practical quantum networks and communications, such as space-based and long-distance metropolitan channels. While we show here spectral modes spaced by 1 THz across 300 nm of bandwidth, dense spectral multiplexing with high generation rates is possible by cascading arrays of microcombs. For example, our AlGaAsOI chip (shown on the inset to Fig. \ref{fig:realfiber}) has $>100$ individual microring resonators, which could each be independently tuned and spectrally aligned to match the 100 GHz telecom ITU grid and pumped with fiber arrays to further enhance the key rates or improve the network user capacity. This could enable $>1,000$ independent users with a single photonic chip or end-to-end multiplexed entanglement key rates exceeding $10$ MHz.

\section{Acknowledgements}

This work was supported by the NSF Quantum Foundry through Q-AMASE-i Program (Award No. DMR-1906325), NSF (Award No. CAREER-2045246), and the National Aeronautics and
Space Administration under Subcontract No. UCSB-01
and Amethyst Research, Inc. We also gratefully acknowledge support from the Cisco Research University Gift Program.

\section{Disclosures}
The authors declare no conflicts of interest.

\section{Data Availability}
The data that support the figures in this paper and other findings of this study are available from the corresponding author on reasonable request.

\bibliography{bibliography.bib}

\renewcommand{\thefigure}{S\arabic{figure}}
\setcounter{figure}{0}

\renewcommand{\thesection}{S\arabic{section}}
\setcounter{section}{0}

\section{Heralded Single Photon Purity Measurement}
\begin{figure}[!htb]
\centering
\includegraphics[width=0.65\columnwidth]{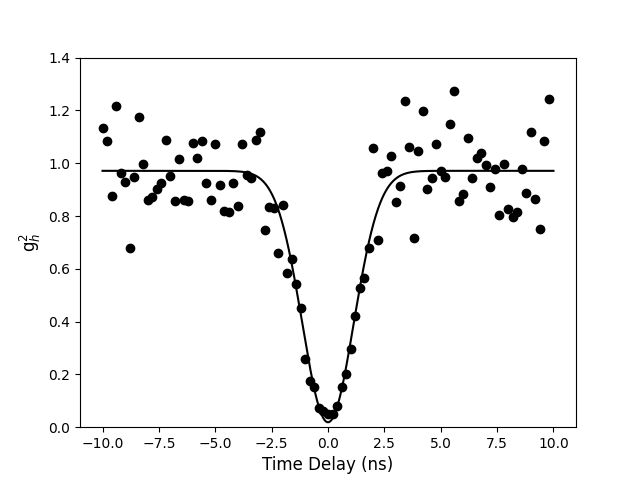}
\caption{Example trace of the heralded g$^2$ for the second adjacent resonance with an on-chip power of 108 $\mu$W. The dip at zero time delay illustrates $>$95$\%$ single photon purity.}
\label{fig:SIg2}
\end{figure}

Verification of single photon purity is completed through a heralded g$^2$ measurement. The signal and idler photon are demultiplexed using DWDMs and sent to separate fibers. The signal photon goes directly to a SNSPD and is used to herald the presence of an idler photon in the other fiber. The idler photon goes through a 3dB splitter which sends half of the light to one detector, and the other half to another detector. The threefold coincidences between the channels are recorded for $>100$ seconds, and the heralded g$^2$ is calculated through $g^2_h=\frac{N_{ABC}N_{A}}{N_{AB}N_{AC}}$ where $N_{ABC}$ is the threefold coincidence count rate, $N_A$ is the heralding single photon count rate, and $N_{AB,AC}$ are the two-fold coincidence counts between the signal channel and each of the two idler channels. Since the nonlinear process generates a single single and idler photon pair (in the low power regime), a dip at zero time delay appears where there are very few threefold coincidence events. The magnitude of this dip relative to the background counts characterizes the single photon purity. Figure \ref{fig:SIg2} illustrates a measurement of 95$\%$ single photon purity.

\section{Visibility Measurement}\label{sectionVis}

\begin{figure}[!htb]
\centering
\includegraphics[width=\columnwidth]{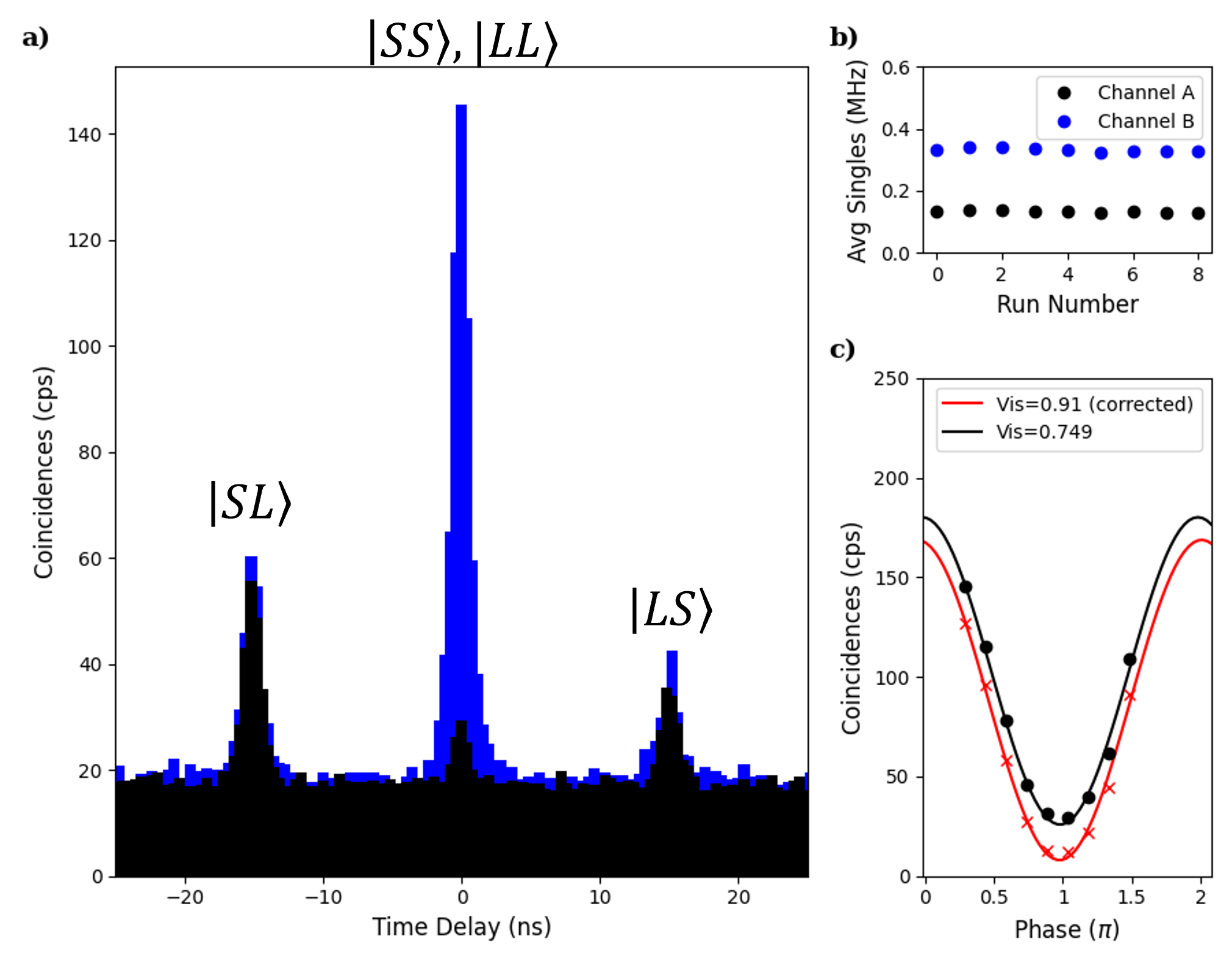}
\caption{a) Coincidence histogram of the two-photon interference at $0.4\pi$ (blue) and $\pi$ (black) relative phase between the long and the short path. b) Single photon flux as the relative phase is swept showing stable flux on both channels. c) Coincidence counts in the central peak as a function of the relative phase along with a sinusoidal fit of the data. The fit shows 74.9$\%$ uncorrected two-photon visibility and 91$\%$ visibility with background subtraction.}
\label{fig:SIVisData}
\end{figure}

To monitor the two-photon interference visibility, a folded Franson interferometer is used. Both the signal and idler photon enter into an unbalanced interferometer with a long and a short path. The resulting two-photon state can be expressed as a summation over $\ket{ij}$, where the $i$ ($j$) index is the path the signal (idler) photon travels, with $i,j$ = [S,L]. The side peaks arise from photons travelling along the $\ket{LS}$ or $\ket{SL}$ paths and are offset from zero by the difference in the long and short path lengths. The central peak at zero delay is due to both photons taking the same paths, $\ket{SS}$ or $\ket{LL}$. Because these states are indistinguishable, the two-photon state is expressed as $\frac{1}{\sqrt{2}}\left(\ket{SS} + exp(i\phi_{i+s})\ket{LL}\right)$. A voltage-controlled phase shifter is used to sweep the relative phase $\phi_{i+s}$ between the two paths to show interference between the states. Fig. \ref{fig:SIVisData}a) shows the difference between the coincidence peaks when a 0.4$\pi$ phase is applied (blue trace) and $\pi$ phase (black trace). The central peak shows two-photon interference visibility of 74.9$\%$ without background subtraction, and 91$\%$ after removing the background counts. This visibility measurement was taken from the second resonance when pumped at 170 $\mu$W.

\begin{figure}[!htb]
\centering
\includegraphics[width=0.65\columnwidth]{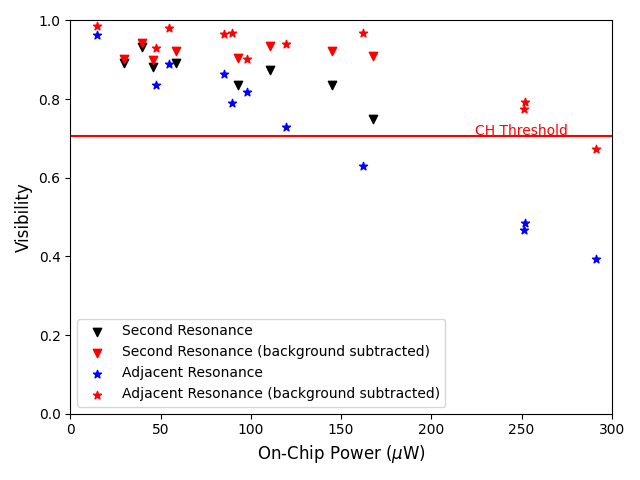}
\caption{Visibility as a function of power for the adjacent and second nearest resonance with and without background subtraction. The red data points indicate the background subtracted results while the blue and black data points indicate the same data as shown in Figure 3d).}
\label{fig:SIVisBack}
\end{figure}

As demonstrated in Figure \ref{fig:SIVisData}c), there is a significant improvement in the visibility after subtracting background coincidence counts. Figure \ref{fig:SIVisBack} illustrates the same plot from Figure 3d) now including both uncorrected (black, blue) and background subtracted data points (red). The visibility still degrades as the power increases, which is expected since the multi-photon pair emission probability increases with power, but the background subtracted data shows the ability to surpass the CH threshold at much higher powers compared to the uncorrected data. Since the DWDM filters do not perfectly extinguish the pump at powers $>$100 $\mu$W, it is likely that the pump light is reducing the visibility and with additional pump filtering, the visibility will approach the background subtracted results.
\end{document}